# Direct Visualization of Relativistic Quantum Scars

Zhehao Ge[1,*,†], Anton M. Graf[2], Joonas Keski-Rahkonen[3,4], Sergey Slizovskiy[5,6], Peter Polizogopoulos[1], Takashi Taniguchi[7], Kenji Watanabe[8], Ryan Van Haren[1], David Lederman[1], Vladimir I. Fal'ko[5,6,9], Eric J. Heller[3,4], Jairo Velasco Jr.[1,†]

[1]*Department of Physics, University of California, Santa Cruz, CA 95064, USA*

[2]*Harvard John A. Paulson School of Engineering and Applied Sciences, Harvard University, Cambridge, MA 02138, USA*

[3]*Department of Physics, Harvard University, Cambridge, MA 02138, USA*

[4]*Department of Chemistry and Chemical Biology, Harvard University, Cambridge, MA 02138, USA*

[5]*Department of Physics and Astronomy, University of Manchester, Oxford Road, Manchester, M13 9PL, UK*

[6]*National Graphene Institute, University of Manchester, Booth Street East, Manchester, M13 9PL UK*

[7]*Research Center for Materials Nanoarchitectonics, National Institute for Materials Science, 1-1 Namiki, Tsukuba 305-0044, Japan*

[8]*Research Center for Electronic and Optical Materials, National Institute for Materials Science, 1-1 Namiki, Tsukuba 305-0044, Japan*

[9]*Henry Royce Institute for Advanced Materials, Manchester, M13 9PL, UK*

*Present Address: Department of Physics, University of California, Berkeley, CA 94720, USA

[†]Email: zge2@ucsc.edu, jvelasc5@ucsc.edu

## Abstract:

Quantum scars refer to eigenstates with enhanced probability density along unstable classical periodic orbits (POs). First predicted 40 years ago[1], scars are special eigenstates that counterintuitively defy ergodicity in quantum systems whose classical counterpart is chaotic[2,3]. Despite the importance and long history of scars, their direct visualization in *quantum* systems remains an open field[4-11]. Here we demonstrate that, by using an *in-situ* graphene quantum dot (GQD) creation and wavefunction mapping technique[12,13], quantum scars are imaged for Dirac electrons with nanometer spatial resolution and meV energy resolution with a scanning tunneling microscope. Specifically, we find enhanced probability densities in the form of lemniscate(∞)-shaped and streak-like patterns within our stadium-shaped GQDs. Both features show equal energy interval recurrence, consistent with predictions for relativistic quantum scars[14,15]. By combining classical and quantum simulations, we demonstrate that the observed patterns correspond to two unstable POs that exist in our stadium-shaped GQD, thus proving they are both quantum scars. In addition to providing the first unequivocal visual evidence of quantum scarring, our work offers insight into the quantum-classical correspondence in relativistic chaotic quantum systems and paves the way to experimental investigation of other recently proposed scarring species such as perturbation-induced scars[16-19], chiral scars[20-22], and antiscarring[23].

Chaos manifests itself ubiquitously in our classical world, causing havoc in various facets of our everyday lives such as long-term weather forecasts and financial market predictions. Still, the quantum nature of chaos remains puzzling. This field of research, known as quantum chaology[24], was first highlighted by Albert Einstein in the early days of quantum mechanics[25,26] and continues to be an active field of research[14-23,27-30]. Essentially, quantum effects tame classical chaos to some extent due to the linearity of the Schrödinger equation, uncertainty principle, and interference[2,31]. One remarkable example of this suppression is demonstrated by quantum scars[1], which refer to eigenstates with enhanced quantum probability density in the vicinity of unstable classical periodic orbits (POs). This phenomenon is now established as a common feature for quantum systems with corresponding chaotic classical dynamics[2,3]. Apart from fundamental appeals to subjects such as quantum ergodicity[32-34] and random matrix theory[35], quantum scars also offer the opportunity to leverage quantum coherence to gain control over broader quantum systems for applications[17,18]. Moreover, the quantum scars discussed here (Heller type[1]) are also recently found to be related to the hotly investigated many-body scars[28-30], which are of current interest because of their potential utility for quantum information.

Since the initial prediction of quantum scars[1], extensive experimental attempts have been carried out to image these states. Although scar analogues have been visualized in various classical wave experiments such as in microwave cavities[36,37], acoustic cavities[38], and fluid surface waves[39], the direct visualization of this phenomenon in real quantum systems remains an open field[4-11]. Despite the slow progress on the experimental frontier, many new types of quantum scars have been proposed in recent years, such as relativistic scars[14,15], perturbation-induced scars[16-19], chiral scars[20-22], and antiscarring[23]. This growing field of quantum scars requires experimental techniques that can directly visualize these states with both high spatial and energy resolution.

To date, most experimental attempts for the imaging of quantum scars have employed scanning gate microscopy in quantum dots (QDs) with corresponding chaotic classical dynamics[5-10]. In these works, features that could be potentially linked to quantum scars were identified. Yet, because of the low resolution, indirect nature of acquiring local density of states (LDOS) and the rough etched QD boundaries in these experiments, whether these features truly correspond to quantum scars remains unclear. Comparing to scanning gate microscopy, scanning tunneling microscope (STM) is a more suitable experimental tool for the visualization of quantum scars because it directly acquires LDOS maps, and such mapping can be obtained with atomic-scale spatial resolution and meV energy resolution. Notably, these capabilities of STM are not possible with scanning gate microscopy. While the 1990-era's quantum corrals[4,40,41] offered a good opportunity for the imaging of quantum scars with STM, the strong wave absorption of the corral walls made such attempts unsuccessful[4]. More recently, electrostatically defined graphene QDs (GQDs)[11-13,42-46] offered a new venue for the imaging of quantum scars with STM. However, earlier attempts have not succeeded so far due to the shallow and smooth confinement potential well of those studied GQDs[11].

In this work, we use STM to directly visualize quantum scars in stadium-shaped GQDs with unprecedented sharp potential wells. In contrast to previously studied circular GQDs[12,13,42-52], which are integrable and have regular (non-chaotic) corresponding classical dynamics as schematized in Fig. 1a, stadium-shaped GQDs are non-integrable and host chaotic classical dynamics (Fig. 1b), making them a suitable platform for studying quantum scars[14,15]. Figure 1c shows the schematic of our experimental set up. An electrostatically defined stadium-shaped GQD is created *in-situ* by a recently developed STM tip bias voltage pulsing technique[12,13] (details in SI section S1). Gate voltage ($V_G$) and tip-sample bias voltage ($V_S$) dependent $dI/dV_S$ mappings are performed to image the probability density of quantum states in the fabricated GQD.

We start by characterizing the shape and potential profile of our GQDs. Figure 1d shows a typical $dI/dV_S$ map taken on the surface of a graphene/hexagonal boron nitride (hBN) heterostructure after the GQD fabrication process. Evidently, an approximately stadium-shaped boundary as outlined by a red dashed line can be seen, which also encloses bright fringes of various wavelengths. We next determine the potential profile of this structure by performing spatially resolved $dI/dV_S$ spectra to track the spatial evolution of the graphene Dirac point. Figures 1e,f show the typical tunneling spectra measured along the horizontal and vertical directions of our stadium-shaped GQD across its center. Here we show $d^2I/dV_S^2(V_S, d)$ plot to help visualize the position of graphene Dirac point (marked by black dots in Figs. 1e,f), raw $dI/dV_S(V_S, d)$ data can be found in SI section S2. We emphasize that our *in-situ* created GQD has a relatively flat potential along the horizontal direction near the stadium center, but along the vertical direction the potential profile is close to a parabolic shape. This type of potential profile is unprecedented for GQDs.

After resolving the QD shape and potential profile of our *in-situ* created GQD, we perform $dI/dV_S$ mappings with different $V_G$ and $V_S$ configurations to search for signatures of quantum scars. Figures 2a-d show $dI/dV_S$ maps measured at different $V_G$ but at a constant bias voltage $V_S = -18$ mV. In general, we see the effective size of the GQD increases with decreasing $V_G$. This behavior stems from the tuning of the GQD potential well sharpness and the GQD Fermi level by $V_G$[11] (see additional $dI/dV_S(V_S, d)$ data at other $V_G$ in SI section S3). In addition, features with enhanced $dI/dV_S$ intensity along closed trajectories can be observed within the GQD across maps at different $V_G$. For example, the red and green boxes in Figs. 2a,b enclose a lemniscate(∞)-shaped and streak-like pattern that mimics previously predicted bow-tie scar and bouncing-ball scar in stadium billiards[1,53].

To better understand the features seen in our $dI/dV_S$ maps, we perform $dI/dV_S$ mappings at a constant $V_G$ with varying $V_S$ instead. Compared to $dI/dV_S$ maps shown in Figs. 2a-d, $dI/dV_S$ maps measured at constant $V_G$ maintain the GQD doping and potential well unchanged, thus these maps more straightforwardly reveal the energy dependence of GQD wavefunctions. Figures 2e-p present $dI/dV_S$ maps measured at a constant $V_G = -19$ V with different $V_S$. At $V_S = -18$ mV, enhanced $dI/dV_S$ amplitude is once again observed along a ∞-shaped trajectory (Fig. 2e). Upon increasing $V_S$, this feature diminishes (Fig. 2f) and reverses to a suppressed $dI/dV_S$ intensity (Fig. 2g) before an enhancement reappears (Fig. 2h). A similar behavior is observed for the enhanced $dI/dV_S$ amplitude along a streak-like trajectory in our $dI/dV_S$ maps. First, an enhancement is observed at $V_S = 6$ mV (Fig. 2k), then it gradually fades (Figs. 2l,m) and eventually inverts into a suppression (Fig. 2n). Finally, this pattern reemerges (Fig. 2o) until a strong enhancement like that at $V_S = 6$ mV is observed again (Fig. 2p).

By performing $dI/dV_S$ map measurements at several bias voltages and with $V_G = -19$ V, we find the enhanced $dI/dV_S$ intensity along the ∞-shaped and streak-like trajectories reoccur with specific $V_S$ values. As shown in Figs. 3a-h, the enhancement of $dI/dV_S$ intensity along the ∞-shaped trajectory recurs every 6 meV (Fig. 3a-d), while for the streak-like trajectory the recurrence is every 10 meV (Fig. 3e-h). $dI/dV_S$ maps with $V_S$ values in between the ones shown in Figs. 3a-h can be found in SI section S4. Notably, such equal energy interval recurrence is expected for relativistic quantum scars[14,15]. This prediction matches appropriately with our recurrence energy findings because we used monolayer graphene (MLG) as our QD canvas, which hosts relativistic massless Dirac fermions[54].

More quantitatively, the recurrence energy $\Delta E$ for relativistic quantum scars is related to the length $L$ of their corresponding classical periodic orbits as[14] $\Delta E = hv_F/L$, where $h$ is Planck's constant and $v_F \approx 10^6$ m/s is MLG's Fermi velocity. With the high spatial resolution provided by our $dI/dV_S$ maps, we can directly measure $L$ of the observed ∞-shaped and streak-like trajectories by fitting them with short straight-line segments as shown in Figs. 3a and 3e, respectively. We extracted an $L$ of ~ 670 nm and 400 nm, yielding an $\Delta E$ of ~6.2 meV and 10.3 meV for the ∞-shaped and streak-like patterns. These values are both in good agreement with our experimentally observed $\Delta E$. A quantitative connection between $\Delta E$ and $L$ for both the ∞-shaped and streak-like patterns is also observed at other $V_G$ (see SI section S5). Therefore, the observed imprints of POs in our $dI/dV_S$ maps satisfy the hallmark of relativistic quantum scars[14,15].

Apart from $dI/dV_S$ maps, the equal energy interval recurrence of the observed ∞-shaped and streak-like patterns are also revealed in spatially resolved $dI/dV_S$ spectra. Figures 3i,j show $dI/dV_S(V_S, d)$ plots measured at $V_G = -19$ V along a horizontal line across the stadium center and 30 nm off the center as indicated by the red and green lines in Fig 3c, respectively. The existence of the ∞-shaped and streak-like patterns is unclear in the raw $dI/dV_S(V_S, d)$ data shown in Figs 3i,j but can be unraveled by performing fast Fourier transforms (FFTs) of the $dI/dV_S(V_S, d)$ data with respect to $V_S$. The resulting FFT amplitude plots are shown in Figs 3k,l. FFT peaks that correspond to the ∞-shaped and streak-like patterns are indicated by blue and red arrows, respectively. The correspondence between the FFT peaks and $dI/dV_S$ map patterns can be seen by comparing the spatial locations of the FFT peaks and the intersection points (blue dots in Figs. 3c,g) between the tunneling spectra measurement lines (red and green lines in Figs. 3c,g) and the ∞-shaped and streak-like patterns. For example, the $y = 0$ nm and $y = 30$ nm lines have three and two intersection points with the ∞-shaped pattern, which agrees with the number of FFT peaks appear in Figs. 3k,l that correspond to the ∞-shaped pattern. Moreover, Figures 3m,n show the averaged FFT amplitude profile between $d = -20$ nm and 20 nm for Figs. 3k,l, respectively. The FFT peaks that correspond to the streak-like and ∞-shaped patterns appear at ~0.10 meV$^{-1}$ and ~0.17 meV$^{-1}$. This coincides well with the recurrence energies of ~10 meV and ~6 meV for the streak-like and ∞-shaped patterns extracted from the $dI/dV_S$ map data. Similar FFT analyses at other $V_G$ values can be found in SI section S6.

To investigate the origin of our experimentally observed ∞-shaped and streak-like $dI/dV_S$ map patterns more deeply, we carry out classical and quantum-mechanical simulations for a relativistic stadium-shaped QD, which captures the dynamics of our GQD. The QD potential well used in our simulations is extracted from the experimental $dI/dV_S(V_S, d)$ data shown in Figs. 1e,f (more details in SI section S7).

We first discuss the corresponding classical dynamics of our stadium-shaped GQD (methods in SI section S8). Figures 4a,b show the calculated Poincaré surface of section (PSS) for our stadium-shaped GQD at an energy relevant to our experiments. A PSS is a commonly used tool to characterize the dynamics of a classical dynamical system[3]. Here the PSS records a particle's $x$ coordinate and momentum $p_x$ every time the particle crosses the $y = 0$ line when moving inside our stadium-shaped GQD in a chosen direction. Evidently, our stadium-shaped GQD has mixed corresponding classical dynamics (i.e., consisting of both chaotic and regular dynamics), which is indicated by the existence of both a chaotic sea (dark regions) and Kolmogorov-Arnold-Moser (KAM) islands (closed contours) in the PSS[3]. We found the relatively soft confinement of our stadium-shaped GQD leads to the formation of KAM islands, resulting in this mixed behavior (see SI section S9). However, the relatively small size of the KAM islands compared to the chaotic regions suggests that the system is only slightly perturbed from a fully chaotic stadium billiard with hard walls.

Upon closer examination of our classical simulations, we have identified streak-like (Fig. 4c) and ∞-shaped (Fig. 4d) POs that closely mirror the ∞-shaped and streak-like $dI/dV_S$ map patterns observed in our experiments. The stability of these POs is highlighted by their positions in our PSS: the ∞-shaped PO is within one KAM island (blue dot in Fig. 4b) and the streak-like PO is in the chaotic sea (orange dot in Fig. 4a). While additional POs exist (see SI section S10), they were not observed in our experiments possibly due to longer coherence requirements. Notably, our classical analysis suggests that the enhanced streak-like amplitude observed in our experiments corresponds to a quantum scar, known better as a bouncing ball scar[53]. Interestingly, the existence of this state is surprising because Dirac fermions hitting a graphene pn junction with normal incidence should have 100% transmission probability due to Klein tunneling[55]. Additional future studies are needed to address this apparent disagreement with Klein tunneling. The enhanced amplitude of the ∞-shaped pattern spatially coincides with a stable classical PO, but it also corresponds to a quantum scar in our case. Due to the uncertainty principle, quantum dynamics are less affected by phase space features smaller than Planck's constant; hence, as the area of the corresponding KAM island is smaller than Planck's constant (Fig. 4b), the ∞-shaped orbit can be interpreted as quantum mechanically unstable[31]. Consequently, the ∞-shaped pattern also corresponds to a quantum scar.

To gain deeper insights into the correspondence between quantum and classical dynamics, we next perform quantum dynamics simulations for our stadium-shaped GQD (method in SI section S11). A wavepacket analysis is used to reveal the existence of scarred states with equal energy intervals that are governed by specific unstable POs. This technique has been previously established and demonstrated in a non-relativistic hard-wall stadium[1]. By preparing a test wavepacket to have a large overlap with a given PO and propagating it in time, we can identify eigenstates that display quantum scarring associated with that particular PO. Figures 4e,f showcase the power spectral density (PSD) of wavepackets initialized along two distinct periodic orbits: the ∞-shaped and streak-like POs. This decomposition of the wavepackets serves as "scarmometer" providing valuable information about the contribution of specific scarred eigenstates that lead to the surprisingly long-term fidelity recurrences observed in the evolution of the wavepackets. The energy of each PSD peak in Figs 4e,f corresponds to the eigenenergy of eigenstates scarred by the ∞-shaped and streak-like POs, respectively. Notably, both ∞-shaped (Fig. 4e) and streak-like (Fig. 4f) quantum scars reappear with equal energy interval, in agreement with the behavior of relativistic quantum scars[14,15] and our experimental findings. In addition, the simulated recurrence energy for both ∞-shaped (~6 meV) and streak-like (~10 meV) quantum scars are close to our experimentally observed $\Delta E$. Finally, the experimentally observed ∞-shaped and streak-like patterns are also found in related TB simulations (SI section S12). Consistent with the aforementioned analysis, our TB simulations further validate the equal energy interval recurrence of quantum scars in a stadium-shaped GQD, aligning with the prediction of relativistic quantum scars[14,15].

In summary, our work provides the first unequivocal direct visualization of quantum scars and paves the way towards imaging other recently proposed types of quantum scars, such as chiral scars[20-22] and perturbation-induced scars[16-19]. Chiral scars are of fundamental interest because they are proposed to manifest in confined neutrinos, which can host spontaneous time reversal symmetry breaking[56]. Chaotic graphene QDs discussed in this work provides a viable platform to emulate this relativistic particle physics phenomenon. On the other hand, perturbation-induced scars can enable guiding and steering of electrons across nanoscale transistors[17,18]. This offers a new pathway for quantum control that leverages chaos for novel nanoelectronic devices.

## Methods

**Sample fabrication.** The graphene/hBN sample used in this study was assembled with a standard polymer-based transfer method[57]. A graphene flake exfoliated on a methyl methacrylate (MMA) substrate was mechanically placed on top of an ~20 nm thick hBN flake that rests on a 285 nm $SiO_2/Si^{++}$ substrate. The MMA scaffold was dissolved in a subsequent solvent bath. The assembled heterostructure is then annealed in forming gas ($Ar/H_2$) for ~12 hours at 400 °C to reduce residual polymer after the heterostructure assembly procedure. Next, an electrical contact to the sample is made by thermally evaporating 7 nm of Cr and 200 nm of Au using a metallic stencil mask. To further improve the sample surface cleanliness, the heterostructure is then mechanically cleaned using an AFM[58], which is done in a glovebox filled with $N_2$ gas. We perform sequential scans in contact mode (setpoint of 0.2 V, scanning speed of ~15 μm/sec, and 1024 × 1024 pixels resolution) to sweep regions of ~ 30 × 30 μm² by a Cypher S AFM with Econo-ESP-Au tips from Oxford Instruments. Finally, the heterostructure is annealed in ultra-high vacuum (UHV) at 400 °C for seven hours before being introduced into the STM chamber.

**STM/STS measurements.** The STM/STS measurements were conducted in UHV with pressures better than $1 \times 10^{-10}$ mbar at 4.8 K in a Createc LT-STM. Electrochemically etched tungsten tips calibrated on Au(111) surface were used in the experiments. The lock-in AC signal frequency used for STS measurements was 704 Hz.

**Acknowledgments:** We thank the Hummingbird Computational Cluster team at UC Santa Cruz for providing computational resources for the numerical TB calculations performed in this work. J.V.J. and Z.G. acknowledges support from the National Science Foundation under award DMR-1753367. J.V.J acknowledges support from the Army Research Office under contract W911NF-17-1-0473 and the Gordon and Betty Moore Foundation award #10.37807/GBMF11569. V.I.F. and S.S. acknowledge support from the European Graphene Flagship Core 3 Project. V.I.F. acknowledges support from Lloyd Register Foundation Nanotechnology Grant, EPSRC grants EP/V007033/1, EP/S030719/1 and EP/N010345/1. K.W. and T.T. acknowledge supports from the JSPS KAKENHI (Grant Numbers 21H05233 and 23H02052) and World Premier International Research Center Initiative (WPI), MEXT, Japan. A.M.G acknowledges support from the Harvard Quantum Initiative. J.K.R. acknowledges support from the Emil Aaltonen Foundation and the Oskar Huttunen Foundation.

**Author contributions:** J.V.J. and Z.G. conceived the work and designed the research strategy. Z.G. fabricated the samples and performed data analysis under J.V.J.'s supervision. Z.G. carried out tunneling spectroscopy measurements with assistance from P.P. and J.V.J.'s supervision. A.M.G. and J.K.R. performed quantum dynamics simulations under E.J.H.'s supervision. Z.G. performed classical dynamics and TB simulations with input from S.S. under V.I.F. and J.V.J.'s supervision. K.W. and T.T. provided hBN crystals. R.V.H. and D.L. provided instrument support. Z.G., J.V.J., J.K.R, A.M.G and E.J.H. wrote the paper. All authors discussed the paper and commented on the manuscript.

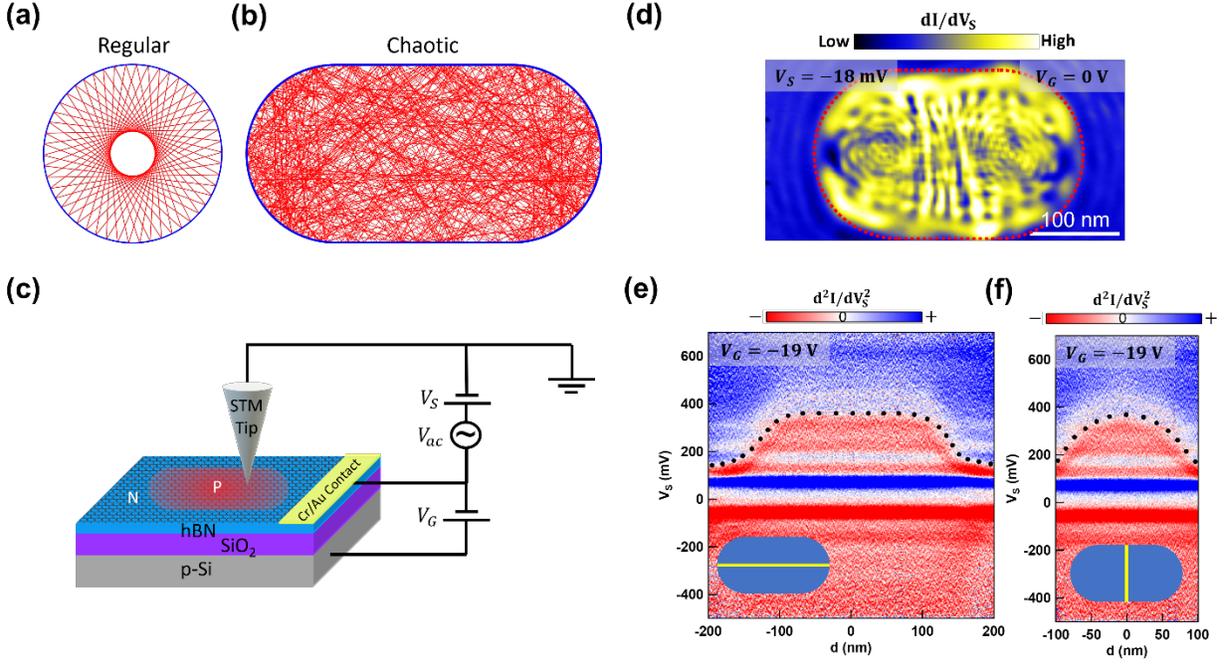

**Figure 1: Scanning tunneling microscope (STM) characterization of an in-situ created stadium-shaped graphene quantum dot (GQD). a-b,** Classical trajectories inside a circular and stadium-shaped billiard. **c,** Schematic of the STM measurement setup. A stadium-shaped GQD defined by a p-n junction is *in-situ* created and can be measured with STM. **d,** Constant bias $dI/dV_S$ map of in-situ created GQD measured at gate voltage $V_G = 0$ V with sample bias voltage $V_S = -18$ mV and a 2 mV ac modulation. The red dashed line indicates the GQD boundary. **e-f,** $d^2I/dV_S^2(V_S, d)$ measured along the horizontal (e) and vertical (f) directions of *in-situ* created stadium-shaped GQD. The measurement direction is indicated by the yellow line in the inset at the bottom of each plot. The black dots indicate the graphene Dirac point position. $d^2I/dV_S^2(V_S, d)$ data are numerically derived from experimentally measured $dI/dV_S(V_S, d)$ data. The set point used to acquire the raw $dI/dV_S(V_S, d)$ data was at a tunneling current $I = 1$ nA, $V_S = -500$ mV with a 10 mV ac modulation.

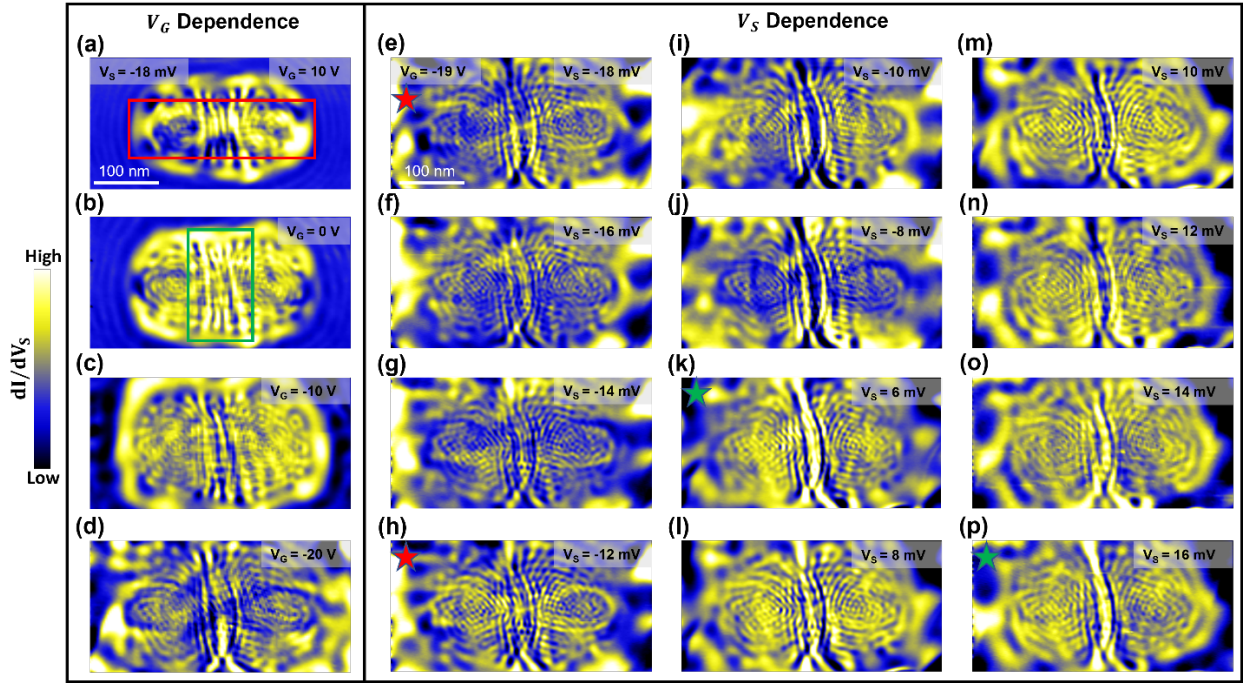

**Figure 2: $V_G$ and $V_S$ dependence of $dI/dV_S$ maps. a-d,** $dI/dV_S$ maps measured at $V_S = -18$ mV with different applied $V_G$ for the same in-situ created GQD shown in Figs. 1d-f. The applied $V_G$ is shown at the top right corner of each $dI/dV_S$ map. A 2 mV ac modulation was used to acquire the $dI/dV_S$ maps. The red box in (a) encloses a ∞-shaped pattern with enhanced $dI/dV_S$ intensity. The green box in (b) encloses a streak-like pattern with enhanced $dI/dV_S$ intensity. **e-p,** $dI/dV_S$ maps measured at $V_G = -19$ V with different applied $V_S$ for the same in-situ created GQD shown in Figs. 1d-f. The applied $V_S$ is shown at the top right corner of each $dI/dV_S$ map. A 2 mV ac modulation was used to acquire the $dI/dV_S$ maps. The red and green stars mark the $dI/dV_S$ maps that show a ∞-shaped and streak-like pattern with enhanced $dI/dV_S$ intensity, respectively.

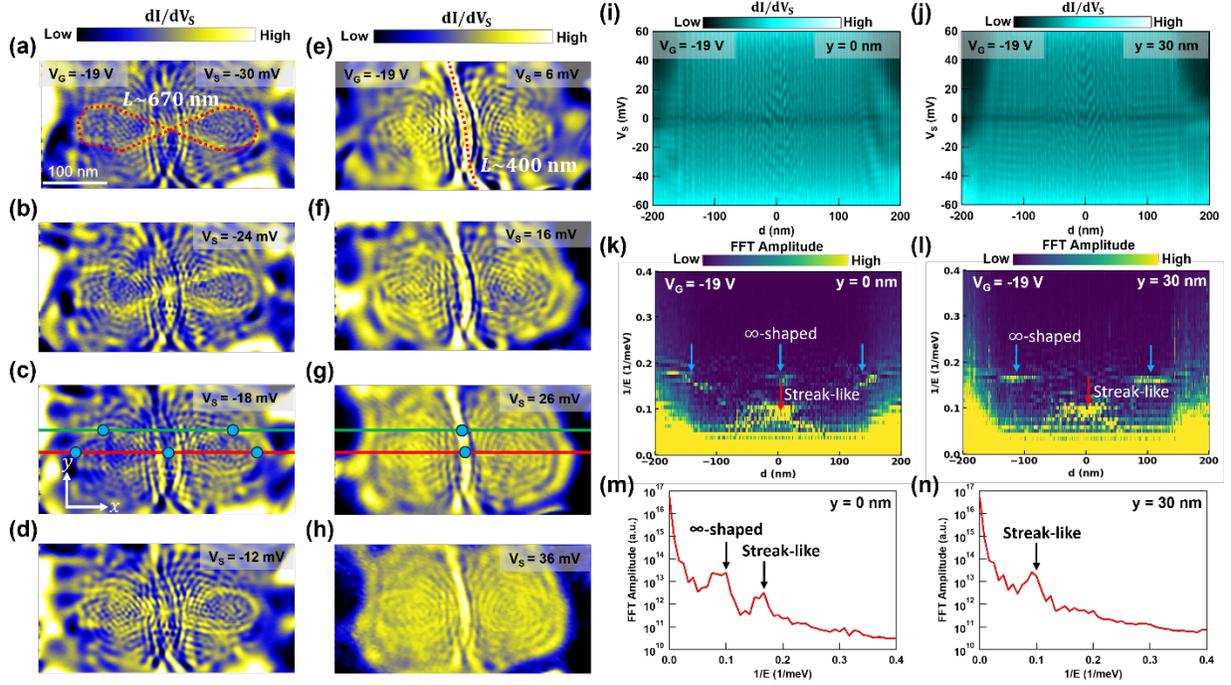

**Figure 3: Equal energy interval recurrence of the ∞-shaped and streak-like $dI/dV_S$ patterns. a-h,** Experimental $dI/dV_S$ maps measured at $V_G = -19$ V with different applied $V_S$ for the same GQD shown in Fig. 2 that show an enhanced $dI/dV_S$ intensity along a ∞-shaped (a) and streak-like (b) trajectories. The applied $V_S$ is shown at the top right corner of each $dI/dV_S$ map. A 2 mV ac modulation was used to acquire the $dI/dV_S$ maps. The red dashed line in (a) and (e) corresponds to the fitted ∞-shaped and streak-like trajectories, respectively. The extracted length $L$ is shown next to the fitted trajectory. The red and green lines in (c) and (g) indicate the positions along which the $dI/dV_S(V_S, d)$ data in (i) and (j) were acquired, respectively. The blue dots in (c) and (d) indicate the intersection points between the red and green lines and the ∞-shaped and streak-like patterns, respectively. **i-j,** $dI/dV_S(V_S, d)$ measured along a horizontal line across the stadium center (i) and 30 nm off the stadium center (j) at $V_G = -19$ V for the same GQD shown in (a)-(h). The setpoint used to acquire the $dI/dV_S$ spectra was $I = 1$ nA, $V_S = -60$ mV with a 2 mV ac modulation. **k-l,** FFT amplitude plot of the $dI/dV_S(V_S, d)$ data shown in (i) and (j), respectively. The blue and red arrows indicate the FFT amplitude peaks that correspond to the ∞-shaped and streak-like patterns, respectively. **m-n,** Averaged FFT amplitude profile between $d = -20$ nm and $d = 20$ nm for the FFT plot shown in (k) and (l), respectively.

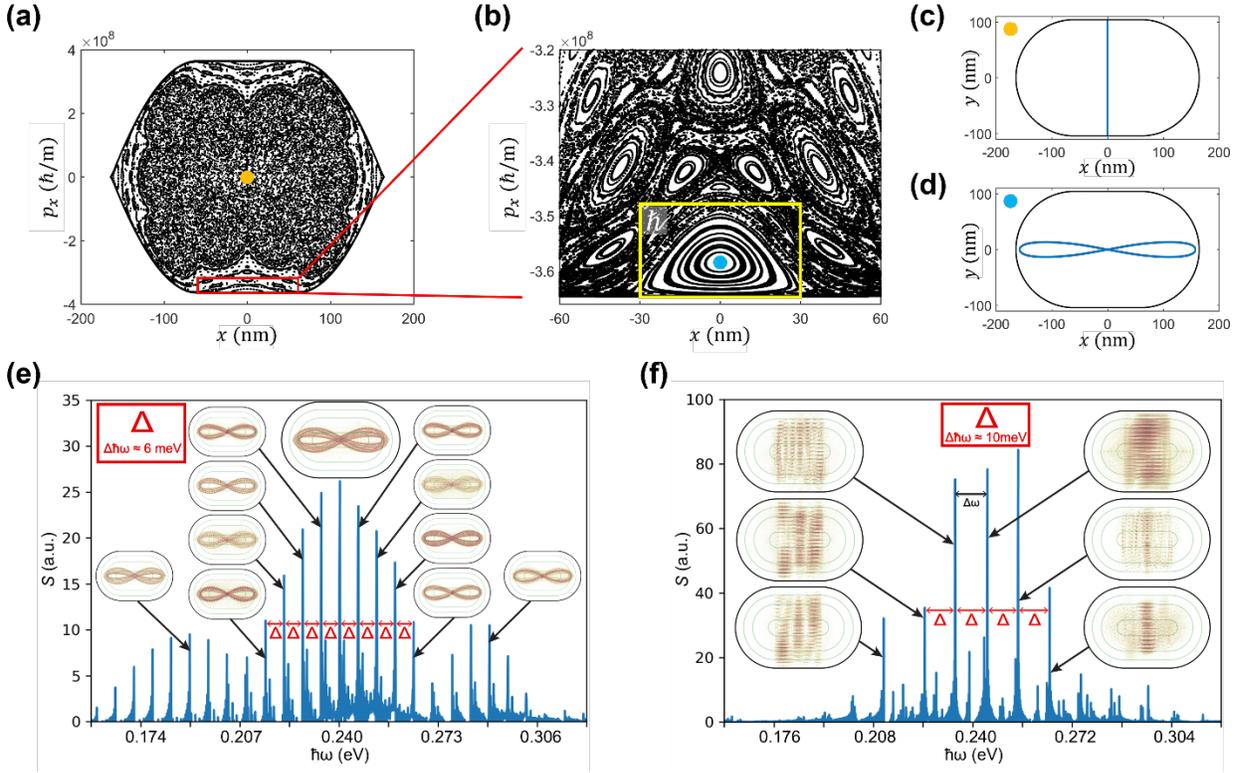

**Figure 4: Classical and quantum dynamics simulations for a stadium-shaped GQD. a-b,** Calculated Poincaré surface of section at 240 meV above the potential well bottom for a stadium-shaped GQD. The yellow box in (b) represents the size of $\hbar$ in phase space. **c-d,** Simulated streak-like (c) and ∞-shaped (d) POs inside a stadium-shaped GQD. The initial conditions used to acquire the POs in (c) and (d) are indicated by the orange dot in (a) and blue dot in (b), respectively. **e-f,** Scarmometers obtained from the Fourier-transformed autocorrelation function of designed wavepackets initialized along ∞-shaped PO (e) and streak-like PO (f), respectively. The insets in (e) and (f) show eigenstates corresponding to the peaks in the spectrum that are scarred by their respective POs. The red arrows indicate the equal energy spacing Δ, a characteristic feature between two adjacent relativistic quantum scars.

# Supplementary Information

# Direct Visualization of Relativistic Quantum Scars

Zhehao Ge[1,*,†], Anton M. Graf[2], Joonas Keski-Rahkonen[3,4], Sergey Slizovskiy[5,6], Peter Polizogopoulos[1], Takashi Taniguchi[7], Kenji Watanabe[8], Ryan Van Haren[1], David Lederman[1], Vladimir I. Fal'ko[5,6,9], Eric J. Heller[3,4], Jairo Velasco Jr.[1,†]

[1]*Department of Physics, University of California, Santa Cruz, CA 95064, USA*

[2]*Harvard John A. Paulson School of Engineering and Applied Sciences, Harvard University, Cambridge, MA 02138, USA*

[3]*Department of Physics, Harvard University, Cambridge, MA 02138, USA*

[4]*Department of Chemistry and Chemical Biology, Harvard University, Cambridge, MA 02138, USA*

[5]*Department of Physics and Astronomy, University of Manchester, Oxford Road, Manchester, M13 9PL, UK*

[6]*National Graphene Institute, University of Manchester, Booth Street East, Manchester, M13 9PL UK*

[7]*Research Center for Materials Nanoarchitectonics, National Institute for Materials Science, 1-1 Namiki, Tsukuba 305-0044, Japan*

[8]*Research Center for Electronic and Optical Materials, National Institute for Materials Science, 1-1 Namiki, Tsukuba 305-0044, Japan*

[9]*Henry Royce Institute for Advanced Materials, Manchester, M13 9PL, UK*

*Present Address: Department of Physics, University of California, Berkeley, CA 94720, USA*

†Email: zge2@ucsc.edu, jvelasc5@ucsc.edu

**Table of Contents**

## S1. Tip pulsing procedure to create stadium-shaped graphene quantum dots (GQDs)

**Tip pulsing procedure A:**

1. Set $V_S = 0.5$ V and $I = 0.5$ nA, then bring $V_G$ to $\sim -60$ V.

2. Open the STM feedback loop.

3. Withdraw the STM tip by $\Delta z \sim 1.1 - 1.2$ nm.

4. Increase $V_s$ to $+5$ V.

5. Wait 2 minutes.

6. Decrease $V_s$ to 0.5 V.

7. Close the STM feedback loop.

**Tip pulsing procedure B:**

1. Set $V_S = 0.5$ V and $I = 0.5$ nA, then bring $V_G$ to $\sim 60$ V.

2. Open the STM feedback loop.

3. Withdraw the STM tip by $\Delta z \sim 1.9 - 2.1$ nm.

4. Increase $V_s$ to $+5$ V.

5. Wait 1 minute.

6. Decrease $V_s$ to 0.5 V.

7. Close the STM feedback loop.

To start creating a stadium shaped GQD, we typically will find a pristine monolayer graphene (MLG) region with at least 400 nm × 200 nm area. After finding a pristine MLG region that meets this requirement, we will first perform tip pulsing procedure A at three locations that are separated by 100 nm as schematized in Fig. S1a. After this step, a large n-doped region that covers the whole identified pristine MLG area can be created.

Next, we will perform tip pulsing procedure B to create smaller p-doped regions in this large n-doped background. More specifically, we will first subsequently perform tip pulsing procedure B at two locations separated by 150 nm around the center of identified pristine MLG region as schematized in Figs. S1b,c. During this process, a single circular p-doped region and a coupled double circular p-doped region can be created subsequently after each tip bias pulse.

After achieving a coupled double circular p-doped region, we will subsequently perform tip pulsing procedure B at another three locations to render the p-doped region to an approximate stadium shape as schematized in Figs. S1d-f. The locations of these three extra tip bias pulses are typically evenly distributed between the centers of the two circular p-doped regions created in the previous step. In the most ideal situation, after completing these three extra tip bias pulses, an approximately stadium shaped GQD as shown in Fig. 1 can be created.

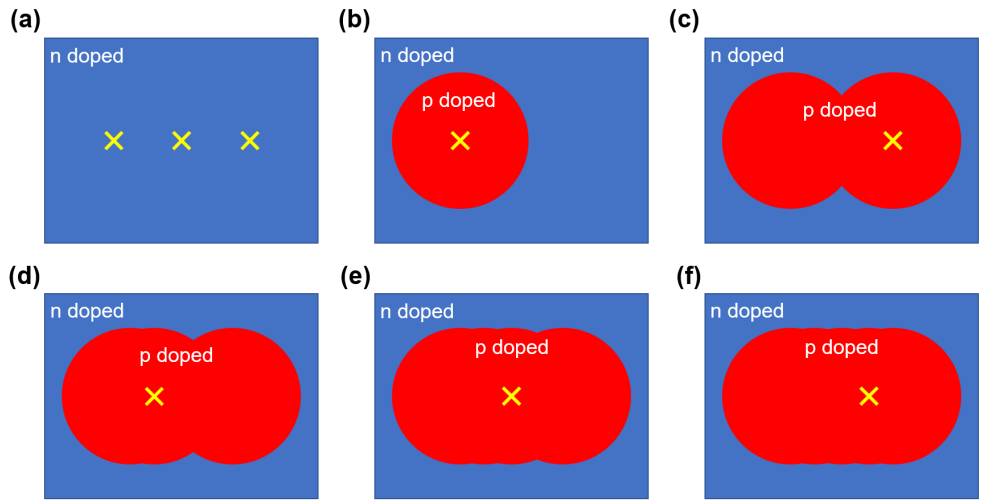

**Figure S1: Schematic of the in-situ creation of a stadium shaped GQD. a-f,** The tip pulsing sequence to create a stadium shaped GQD and the resulting doping pattern after each step. The yellow cross indicates the location of the tip bias pulse. The blue and red colors represent n-doped and p-doped MLG regions, respectively.

## S2. Raw $dI/dV_S(V_S, d)$ data used to get Figs. 1e-f.

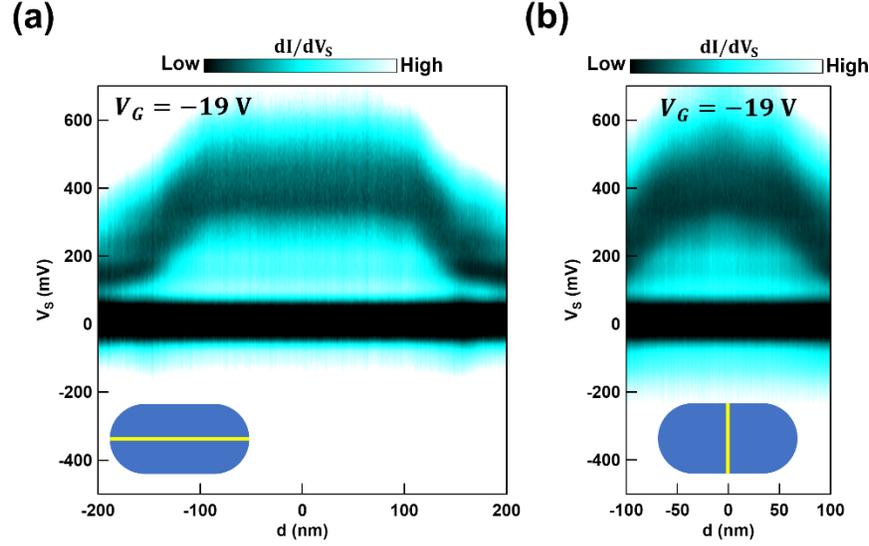

**Figure S2: Raw $dI/dV_S(V_S, d)$ data used to get Figs. 1e-f. a-b,** Experimentally measured $dI/dV_S(V_S, d)$ at $V_G = -19$ V for the same stadium-shaped GQD shown in Fig. 1 across its center along the horizontal (a) and vertical (b) directions. The inset shows the schematic of the measurement direction. The set point used to acquire the $dI/dV_S(V_S, d)$ data was $I = 1$ nA, $V_S = -500$ mV with a 10 mV ac modulation.

## S3. $dI/dV_S(V_S, d)$ data at other $V_G$.

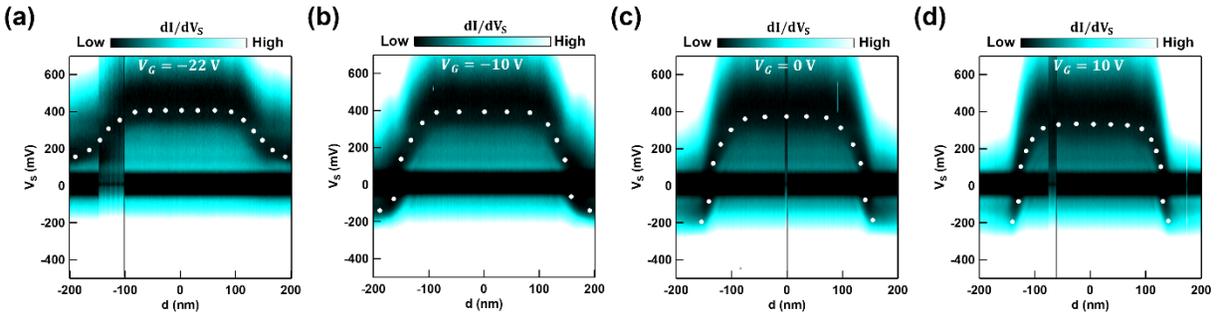

**Figure S3: Gate tunable potential well of in-situ created stadium-shaped GQD. a-d,** Experimentally measured $dI/dV_S(V_S, d)$ at $V_G = -22$ V (a), $-10$ V (b), 0 V (c) and 10 V (d) for the same stadium-shaped GQD shown in Fig. 1 across its center along the horizontal direction. The set point used to acquire the $dI/dV_S(V_S, d)$ data was $I = 1$ nA, $V_S = -500$ mV with a 10 mV ac modulation. The spatial evolution of graphene Dirac point's position is indicated by white dots in (a)-(d). It is evident that the potential well sharpness increases when increasing $V_G$, which results

in smaller stadium sizes at $V_S \sim 0$ V at higher $V_G$. The discontinuity of the tunneling spectra in (a), (c) and (d) is a result of the tip instability during the measurements.

## S4. Extended $dI/dV_S$ map data at $V_G = -19$ V.

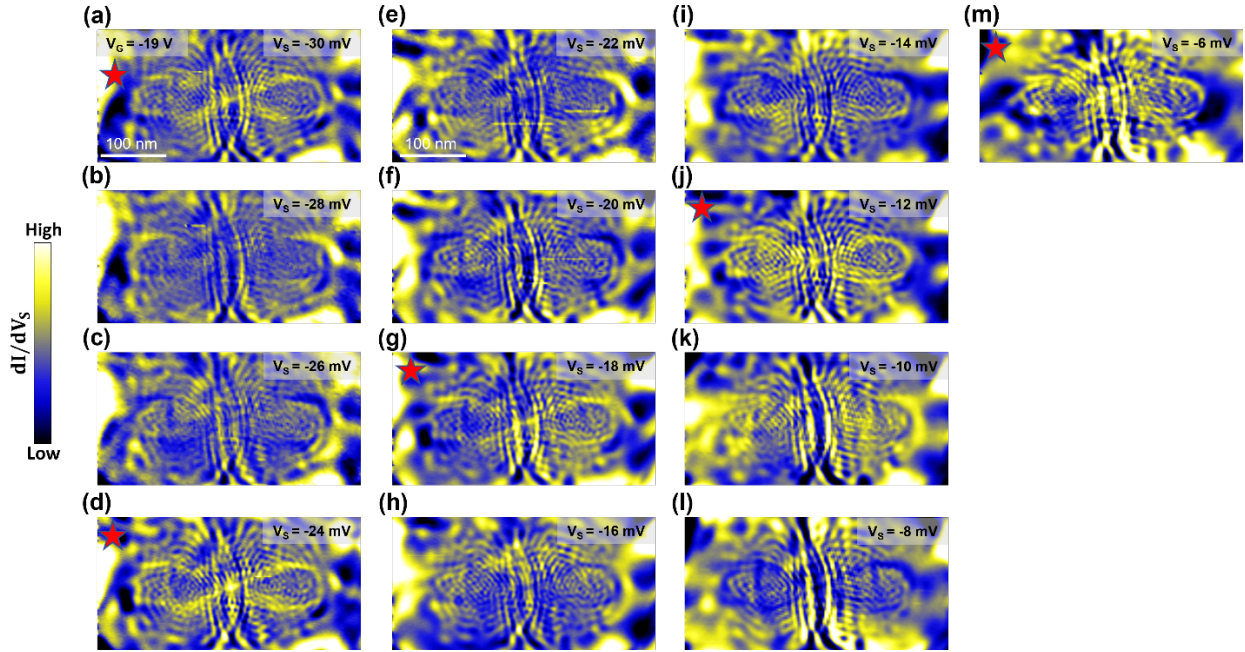

**Figure S4: Extended $dI/dV_S$ map data at $V_G = -19$ V from $V_S = -30$ mV to $-6$ mV. a-m,** $dI/dV_S$ maps measured at $V_G = -19$ V with different applied $V_S$ for the same in-situ created GQD shown in Fig. 3. The applied $V_S$ is shown at the top right corner of each $dI/dV_S$ map. A 2 mV ac modulation was used to acquire the $dI/dV_S$ maps. The red star marks the $dI/dV_S$ maps that show a ∞-shaped pattern with enhanced $dI/dV_S$ intensity.

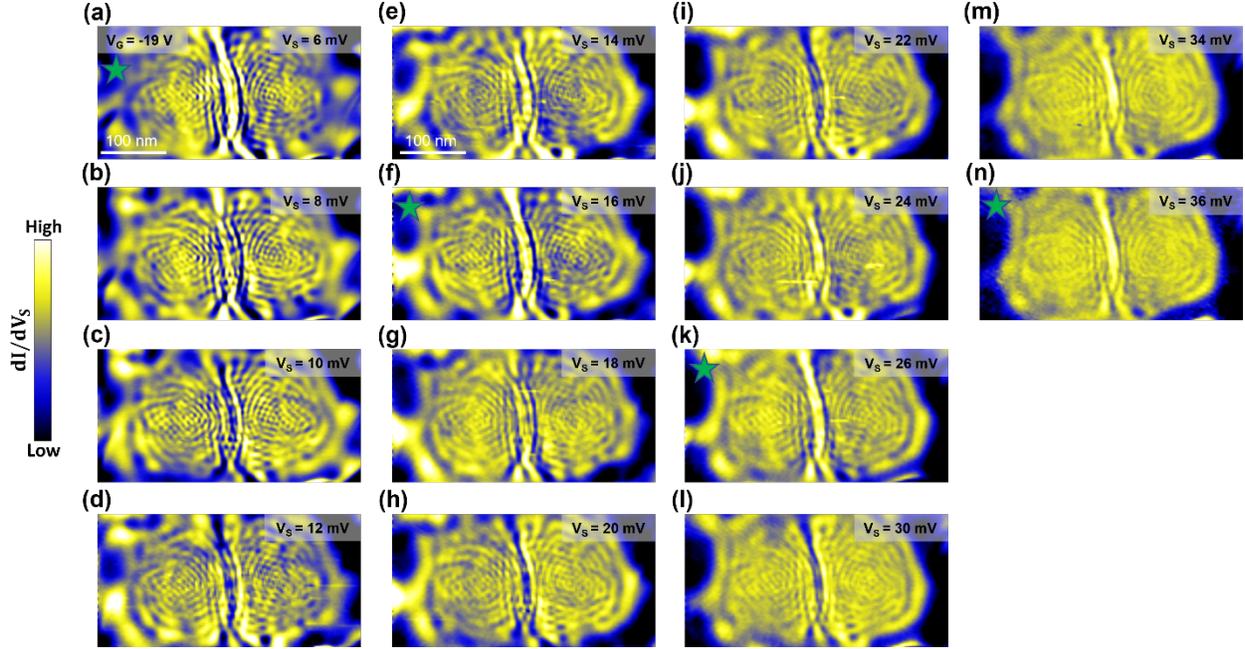

**Figure S5: Extended $dI/dV_S$ map data at $V_G = -19$ V from $V_S = 6$ mV to 36 mV. a-n,** $dI/dV_S$ maps measured at $V_G = -19$ V with different applied $V_S$ for the same in-situ created GQD shown in Fig. 3. The applied $V_S$ is shown at the top right corner of each $dI/dV_S$ map. A 2 mV ac modulation was used to acquire the $dI/dV_S$ maps. The green star marks the $dI/dV_S$ maps that show a streak-like pattern with enhanced $dI/dV_S$ intensity.

## S5. Connection between $\Delta E$ and $L$ for ∞-shaped and streak-like patterns at other $V_G$.

The quantitative connection between $\Delta E$ and $L$ for both the ∞-shaped and streak-like trajectories is also observed at other gate voltages. Fig. S6 shows the experimentally measured $dI/dV_S$ maps that have an enhanced $dI/dV_S$ along a ∞-shaped trajectory at $V_G = -22$ V, 0 V and 10 V. The observed recurrence energy $\Delta E$ is around 6 meV, 8 meV and 8 meV for $V_G = -22$ V, 0 V and 10 V, respectively. The length $L$ of the extracted ∞-shaped trajectory at $V_G = -22$ V, 0 V and 10 V (indicated by red dashed lines in Fig. S6) is around 757 nm, 532 nm and 550 nm, respectively. This leads to a theoretical $\Delta E$ around 5.5 meV, 7.8 meV and 7.5 meV at $V_G = -22$ V, 0 V and 10 V, which is in good agreement with the experimentally observed $\Delta E$ at each $V_G$. Fig. S7 shows the experimentally measured $dI/dV_S$ maps that have an enhanced $dI/dV_S$ along streak-like trajectories at $V_G = -10$ V, 0 V and 10 V. A $\Delta E$ of 12 meV, 14 meV and 16 meV is observed at $V_G = -10$ V, 0 V and 10 V, respectively. By doing the similar analysis as for the ∞-shaped trajectories, we can get the theoretical $\Delta E$ for streak-like trajectories to be around 11.7 meV, 14.4 meV and 16.1 meV at $V_G = -10$ V, 0 V and 10 V, respectively, which is also in good agreement with the experimentally observed $\Delta E$ at each $V_G$.

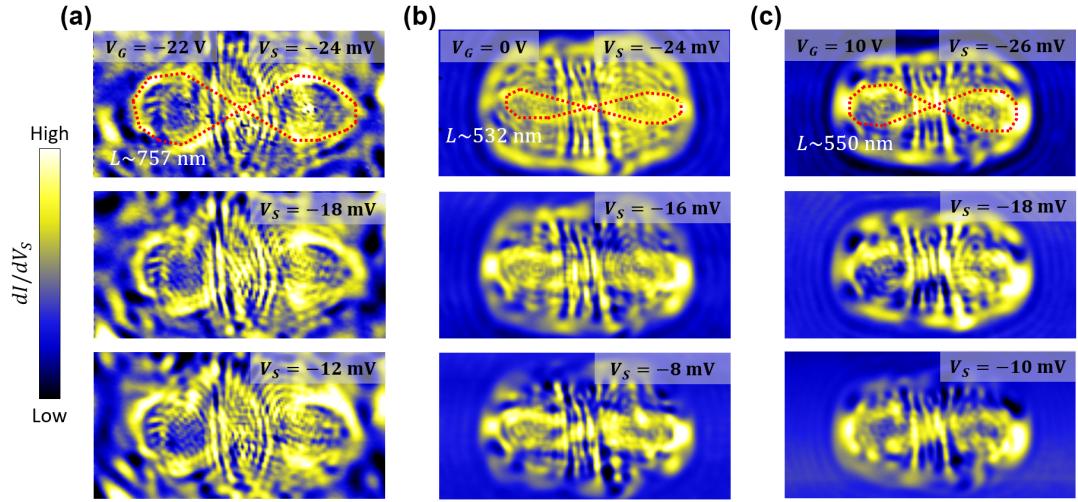

**Figure S6: Connection between $\Delta E$ and $L$ for ∞-shaped trajectories at other $V_G$. a-c,** Experimental $dI/dV_S$ maps at $V_G = -22$ V (a), 0 V (b) and 10 V (c) that show an enhanced $dI/dV_S$ along ∞-shaped trajectories. The corresponding $V_S$ for each map is shown at the top right corner of each map. The red dashed line in the top most map in (a)-(c) corresponds to the fitted ∞-shaped trajectory. The extracted length $L$ is shown next to the fitted trajectory. The scanning window size is 400 nm × 200 nm for all maps.

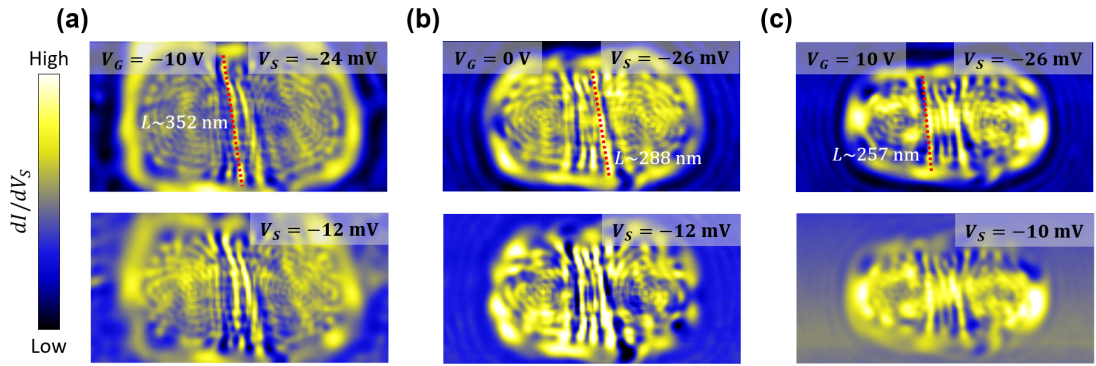

**Figure S7: Connection between $\Delta E$ and $L$ for streak-like trajectories at other $V_G$. a-c,** Experimental $dI/dV_S$ maps at $V_G = -10$ V (a), 0 V (b) and 10 V (c) that show an enhanced $dI/dV_S$ along streak-like trajectories. The corresponding $V_S$ for each map is shown at the top right corner of each map. The red dashed line in the top most map in (a)-(c) corresponds to the fitted

streak-like trajectory. The extracted length $L$ is shown next to the fitted trajectory. The scanning window size is 400 nm × 200 nm for all maps.

## S6. FFT analysis for $dI/dV_S(V_S, d)$ data at other $V_G$.

Fig. S8a-c shows $dI/dV_S(V_S, d)$ plots measured at $V_G = -10$ V, 0 V and 10 V along a horizontal line across the stadium center for the same in-situ created stadium-shaped GQD shown in the main text. Then their corresponding FFT amplitude plots are shown in Figs. S8d-f, respectively, FFT amplitude peaks that correspond to the recurrence energy $\Delta E$ of the ∞-shaped and streak-like patterns are indicated by the blue and red arrows. Figs. S8g-i shows the averaged FFT amplitude profile between $d = -20$ nm and 20 nm for Figs. S8d-f, respectively. The FFT peaks correspond to the ∞-shaped orbit appear at ~0.1417 meV$^{-1}$, ~0.1250 meV$^{-1}$ and ~0.1209 meV$^{-1}$ at $V_G = -10$ V, 0 V and 10 V, respectively. These FFT peak positions correspond to a $\Delta E$ of ~7.1 meV, ~8.0 meV and ~8.3 meV at $V_G = -10$ V, 0 V and 10 V, agreeing well with the experimentally observed $\Delta E$ for ∞-shaped patterns shown in Fig. S6. Then the FFT peaks correspond to the streak-like orbit appear at ~0.0834 meV$^{-1}$, ~0.0709 meV$^{-1}$ and ~0.0625 meV$^{-1}$ at $V_G = -10$ V, 0 V and 10 V, respectively. These FFT peak positions correspond to a $\Delta E$ of ~12.0 meV, ~14.1 meV and ~16.0 meV at $V_G = -10$ V, 0 V and 10 V, also agreeing well with the experimentally observed $\Delta E$ for the streak-like patterns shown in Fig. S7.

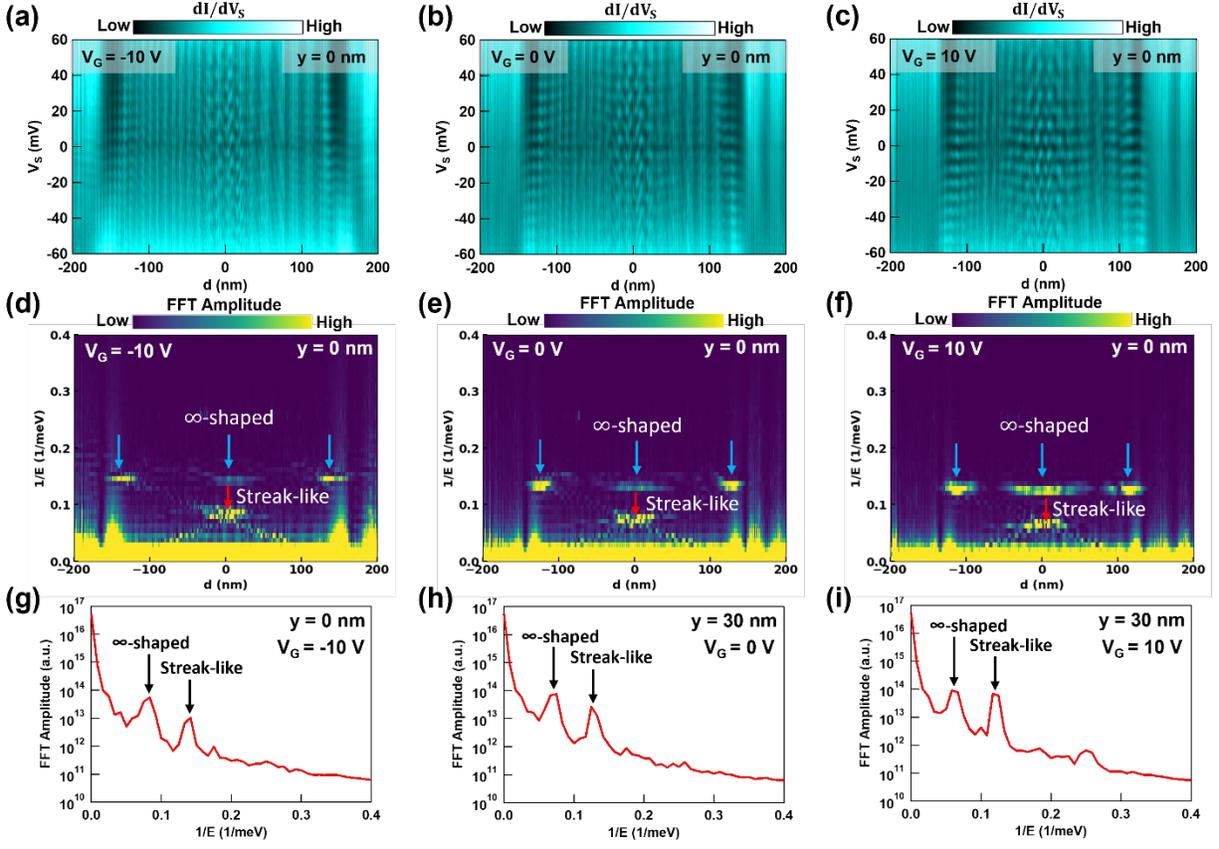

**Figure S8: FFT analysis for $dI/dV_S(V_S, d)$ data at other $V_G$. a-c,** $dI/dV_S(V_S, d)$ measured along a horizontal line across the stadium center at $V_G = -10$ V (a), 0 V (b) and 10 V (c) for the same GQD shown in maintext. The setpoint used to acquire the $dI/dV_S$ spectra was $I = 1$ nA, $V_S = -60$ mV with a 2 mV ac modulation. **d-f,** FFT amplitude plot of the $dI/dV_S(V_S, d)$ data shown in (a)-(c), respectively. The blue and red arrows indicate the FFT amplitude peaks that correspond to the ∞-shaped and streak-like patterns, respectively. **g-i,** Averaged FFT amplitude profile between $d = -20$ nm and $d = 20$ nm for the FFT plot shown in (d)-(f), respectively.

## S7. Simulation potential well.

Figure S9a shows the potential well of the stadium shaped GQD used in our classical, tight-binding (TB) and wave-packet simulations. To compare the similarity between the potential well used in our simulations and the actual potential well of stadium shaped GQD in our experiments, we overlay potential line cuts from Fig. S9a to experimental $d^2I/dV_S^2(V_S, d)$ data taken from the corresponding direction, such comparisons are shown in Figs. S9b-c. The $d^2I/dV_S^2(V_S, d)$ plot shown in Figs. S9b-c are from the same stadium shape GQD shown in Figs. 1e,f at $V_G = -19$ V along the horizontal and vertical directions across the stadium center, respectively. Here we use $d^2I/dV_S^2(V_S, d)$ plot instead of $dI/dV_S(V_S, d)$ plot to better identify the energy position of the local minimum in the $dI/dV_S$ spectra, which corresponds to monolayer graphene's Dirac point. Evidently, in Figs. S9b-c, the line cut of the simulation potential generally is very similar to the experimental potential line cut along both the horizontal (Fig. S9b) and vertical (Fig. S9c) directions.

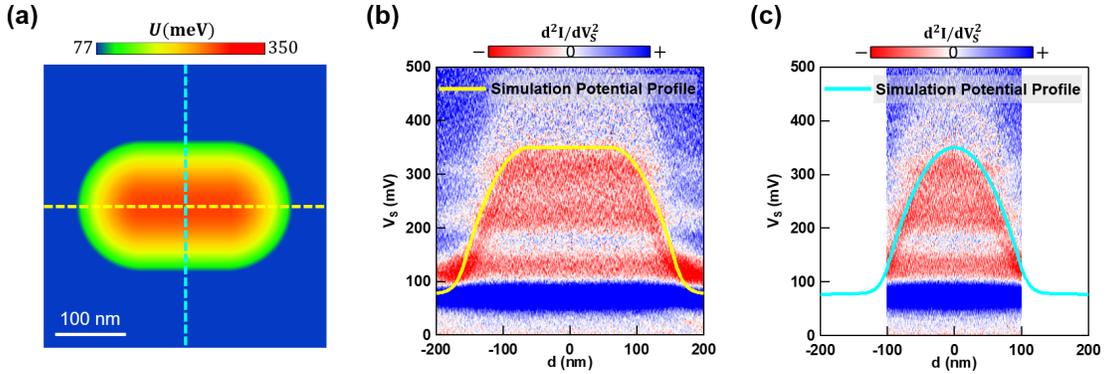

**Figure S9: Comparison between the stadium potential well in simulation and in experiment. a,** Color plot of the potential well used in our classical, tight-binding and wave-packet simulations for a stadium shaped GQD. **b-c,** Zoom-in of the same $d^2I/dV_S^2(V_S, d)$ data shown in Figs. 1e,f. The yellow and blue solid lines in (b) and (c) are the potential line cuts from the potential well shown in (a) along the yellow and blue dashed lines, respectively.

**S8. Classical dynamical simulation for stadium-shaped GQD.**

The classical Hamiltonian for a stadium-shaped GQD can be expressed as

$$H = v_F\sqrt{p_x^2 + p_y^2} + V(x,y)$$

Here $v_F \approx 1.00 \times 10^6$ m/s is MLG's Fermi velocity and $V(x,y)$ is the potential well of the stadium shaped GQD. The potential well shown in Fig. S9a is used in the model. Then the classical orbits and Poincaré surface of section (PSS) for the stadium shaped GQD is simulated by solving Hamilton's equations with Runge-Kutta method.

## S9. Poincaré surface of sections of stadium-shaped GQDs with different potential well sharpness.

Figures S10a-c show the calculated PSS for stadium-shaped GQDs with different potential well sharpness. The potential well sharpness of the stadium-shaped GQD is characterized by parameter $\kappa$, which describes the sharpness of the parabolic like potential well ($U(y) = \kappa y^2$) along the vertical direction across the stadium center as shown in Fig. 1f and Fig. S9c. We notice that, as the potential well sharpness reduces, the size of the Kolmogorov-Arnold-Moser (KAM) islands gradually increases, which means the classical dynamics starting to be less chaotic. In our experiment, the $\kappa$ is around 0.023 meV/nm² for the stadium shown in Fig. 1, which is in between the cases shown in Fig S10b and Fig. S10c.

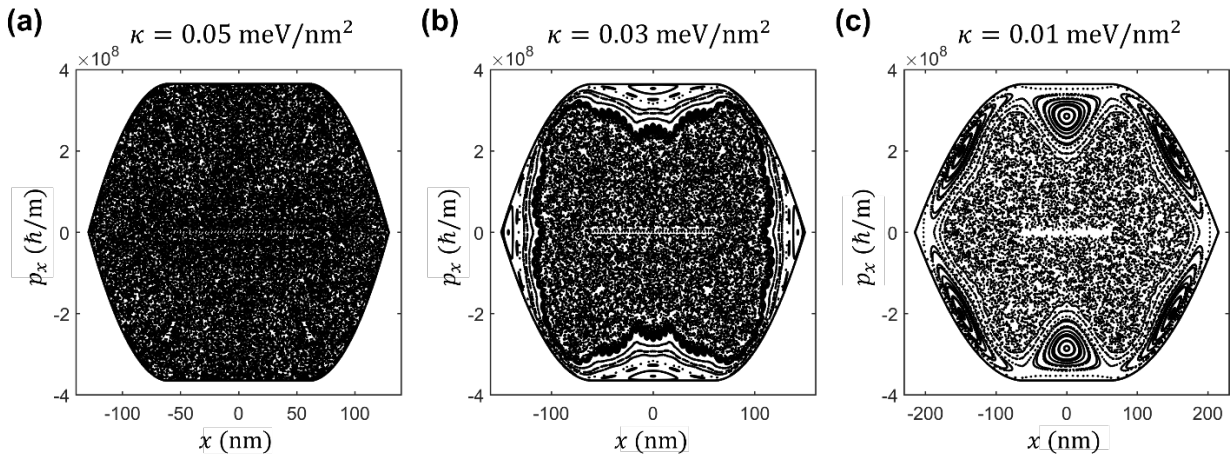

**Figure S10: Impact of potential well sharpness on classical dynamics of stadium-shaped GQDs. a-c,** Simulated PSS for stadium-shaped GQDs with different potential well sharpness. All the PSS are calculated at $E = 240$ meV above the potential well bottom. The value of the $\kappa$ used in the simulation is 0.05 meV/nm², 0.03 meV/nm² and 0.01 meV/nm² for (a)-(c), respectively.

## S10. Additional periodic orbits that exist in stadium-shaped GQD.

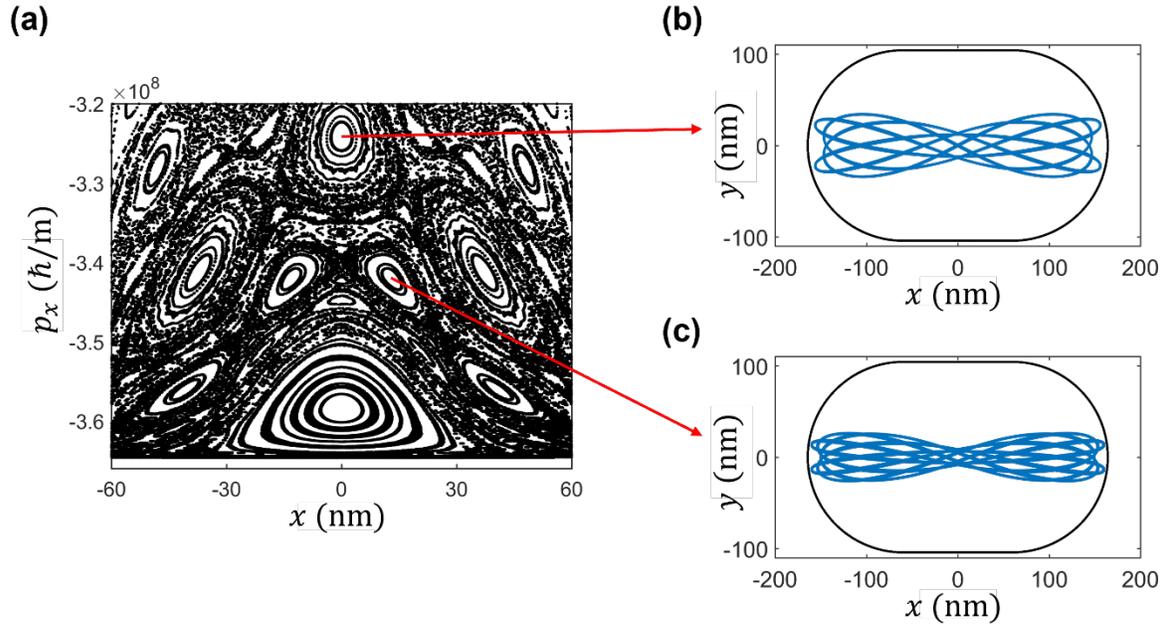

**Figure S11: Additional periodic orbits that exist in stadium-shaped GQD. a,** The same calculated Poincaré surface of section (PSS) for a stadium-shaped GQD as shown in Fig. 4b. **b-c,** Two additional stable periodic orbits for a stadium-shaped GQD that reside in the stable islands of PSS shown in (a).

## S11. Wavepacket simulation for stadium-shaped GQD.

Gaussian wavepackets are commonly employed to investigate eigenstate scarring. When a wavepacket $|\psi\rangle$ is positioned on a scar-generating periodic orbit, its magnitude of the autocorrelation $|\langle\psi(t)|\psi(0)\rangle|$ is large for an eigenstate scarred by the particular periodic orbit, pointing out the utility of wave packets in detecting scarred eigenstates. In the study of the main text, we have employed the following Gaussian wavepacket as a scar detector:

$$\psi(r,0) = \mathcal{N}\exp\left[-\frac{|r-r_0|^2}{2\sigma^2} - \frac{ip_0}{\hbar}(r-r_0)\right],$$

where $\mathcal{N}$ is a normalisation constant. Here, we launch the wavepacket centered at point $r_0$ into a direction $p_0$ determined by a given periodic orbit. For the ∞-shaped scar, we choose $r_0 = (0,0)$, but we slightly move the launching point in the vertical direction enabling a better detection for bouncing-ball scars. However, in both cases, we set to the average energy $\langle E \rangle$ of the wavepacket to match the experimental value of 0.24 eV. Since the stadium potential is approximately flat in the vicinity of the launching point $r_0$, we can set $p_0 = \sqrt{2m\langle E \rangle}$. Furthermore, we can adjust the width $\sigma$ of the wavepacket to improve the scar sensitivity. While keeping the initial momentum $p_0$ as well-defined, we optimize the test Gaussian by squeezing it in the direction perpendicular to the initial momentum, consequently leading to elongation in the direction of the periodic orbit.

The evolution of the test Gaussian is computed via third-order split-operator method for a (pseudo-)relativistic system with a Hamiltonian given below:

$$H = v_F\sqrt{p_x^2 + p_y^2} + V(x,y).$$

Here $v_F \approx 1.00 \times 10^6$ m/s is MLG's Fermi velocity and $V(x,y)$ is the potential well of the stadium shaped GQD. The potential well shown in Fig. S9a is used in the model. The presented scarmometer spectrum $S(\omega)$ is then defined as the time Fourier transform of the autocorrelation function, i.e.,

$$S(\omega) = \frac{1}{2T}\int_{-T}^{T}\exp(-\omega t)\langle\psi(0)|\psi(t)\rangle dt.$$

In the scarmometer, the eigenvalues $\omega_n$ of a system are revealed as peaks with a weight $p_n^2 = |\langle\psi(0)|n\rangle|^2$, where $|n\rangle$ is the eigenstate corresponding to the energy $\omega_n$, or formally we can write

$$\lim_{T\to\infty} S(\omega) = \sum_n p_n^2 \delta(E - E_n).$$

Since the test wavepacket $|\psi(0)\rangle$ is initialized according to a specific classical periodic orbit, the most prominent peaks are associated with eigenstates that significantly overlap with the wavepacket evolution, i.e., a state scarred by the given periodic orbit. Consequently, the larger this weight, the easier it is to numerically extract the (scarred) eigenstate $|n\rangle$.

The scarmometer gives us the (approximately) eigen-energies $\hbar\omega_n$ that can be used to filter out just the corresponding eigenstate $|n\rangle$ in the following way

$$p_n|n\rangle = \lim_{T\to\infty} \frac{1}{2T} \int_{-T}^{T} \exp(-\omega_n t) |\psi(t)\rangle dt,$$

after which we renormalize the resolved state $p_n|n\rangle$ to achieve the sought eigenstate $|n\rangle$. The assumption for this procedure is that the eigen-energies are non-degenerate, which is valid for a chaotic quantum system, such as the stadium-shaped GQD considered in the main text.

**S12. Tight-binding simulation for stadium-shaped GQD.**

The stadium-shaped GQD is modeled by the following tight-binding (TB) Hamiltonian within the first-nearest-neighbor-hopping approximation on a 1.6 um × 1.4 um monolayer graphene mesh:

$$H = \sum_i V(\vec{R}_i^A) a_i^\dagger a_i + \sum_i V(\vec{R}_i^B) b_i^\dagger b_i - \sum_{<i,j>} \gamma_0 (a_i^\dagger b_j + b_j^\dagger a_i)$$

where the operators $a_i^\dagger(a_i)$ and $b_i^\dagger(b_i)$ create (annihilate) an electron on site $\vec{R}_i$ of sublattice $A$ and $B$, respectively. The stadium-shaped GQD is defined by varying the onsite energy $V$ of carbon atoms at the position $\vec{r} = \vec{R}_i$. The $V(\vec{r})$ used for the stadium-shaped GQD is shown in Fig. S9a. The hopping parameter between the nearest carbon atoms is $\gamma_0$, where we used $\gamma_0 = 3.1\ eV$. This corresponds to a Fermi velocity $v_F \approx 1.00 \times 10^6$ m/s for graphene. The local density of states $LDOS(E, r)$ was computed numerically from the above Hamiltonian using the Pybinding package[1], which uses the kernel polynomial method[2] to solve the Hamiltonian.

As shown in Figs. S12, both the ∞-shaped and streak-like patterns observed in our experimental $dI/dV_S$ maps can be reproduced by TB simulated $dI/dV_S$ maps. In addition, our TB simulation also reproduces the equal energy spacing recurrence of the ∞-shaped (Fig. S12a) and streak-like (Fig. S12b) patterns with $\Delta E$ similar to our experimental findings. Furthermore, by overlaying the POs revealed in our classical simulations (Figs. 4c,d) on top of the simulated $dI/dV_S$ maps in Figs. S12a,b (red traces), a good quantum classical correspondence is observed. Given the length $L$ of the simulated classical ∞-shaped PO ($L \approx 651$ nm) and streak-like PO ($L \approx 412$ nm), theoretical recurrence energy $\Delta E$ can be estimated as 6.35 meV and 10.04 meV for the ∞-shaped and streak-like patterns, respectively. These values agree well with the observed $\Delta E$ for the ∞-shaped ($\Delta E \approx 6$ meV) and streak-like patterns ($\Delta E \approx 10$ meV) in our TB simulation results, demonstrating strong quantitative support for the classical to quantum correspondence for relativistic quantum scars.

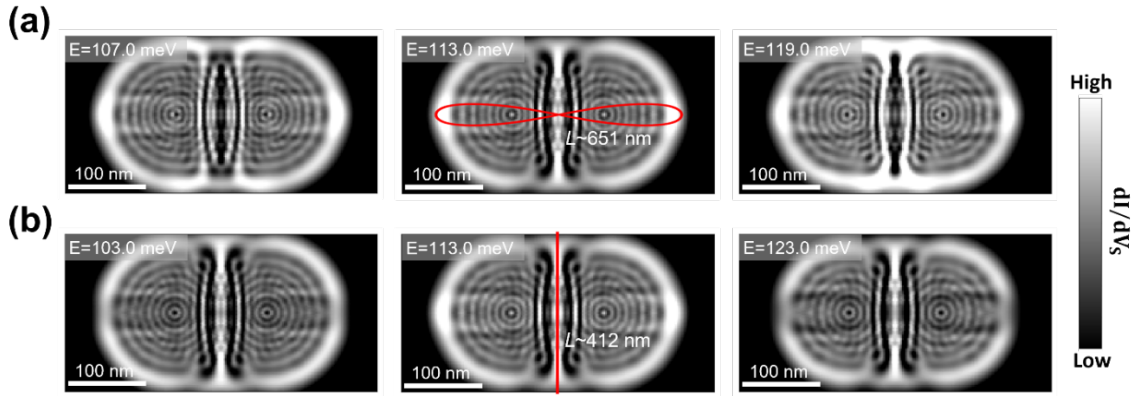

**Figure S12: TB simulated $dI/dV_S$ maps for a stadium-shaped GQD. a-b,** Simulated $dI/dV_S$ maps with enhanced $dI/dV_S$ intensity along the ∞-shaped and streak-like POs, respectively. The energy of each $dI/dV_S$ map is listed at the top left corner of each map. The red lines in (a) and (b) are the calculated corresponding classical ∞-shaped and streak-like POs, respectively. The length of each PO is shown next to them.